\documentclass[doublecol]{epl2}
\usepackage{graphicx}
\usepackage{color}
\usepackage{epsfig}
\usepackage{amssymb}

\begin{document}

\title{Infrared scintillation yield in gaseous and liquid argon}

\author{A. Buzulutskov\footnote{Corresponding author. Email: buzulu@inp.nsk.su}, A. Bondar, A. Grebenuk}

\institute{Budker Institute of Nuclear Physics, Novosibirsk, 630090,
Russia}


\abstract{The study of primary and secondary scintillations in noble
gases and liquids is of paramount importance to rare-event
experiments using noble gas media. In the present work, the
scintillation yield in gaseous and liquid Ar has for the first time
been measured in the near infrared (NIR) and visible region, both
for primary and secondary (proportional) scintillations, using
Geiger-mode avalanche photodiodes (G-APDs) and pulsed X-ray
irradiation.  The primary scintillation yield of the fast component
was measured to be 1.7$\times$10$^4$ photon/MeV in gaseous Ar in the
NIR, in the range of 690-1000 nm, and 5.1$\times$10$^2$ photon/MeV
in liquid Ar, in the range of 400-1000 nm. Proportional NIR
scintillations (electroluminescence) in gaseous Ar have been also
observed; their amplification parameter at 163 K was measured to be
13 photons per drifting electron per kV. Possible applications of
NIR scintillations in high energy physics experiments are
discussed.}

\maketitle

\section{Introduction}

The study of primary and secondary scintillations in noble gases and
liquids is of paramount importance to rare-event experiments using
noble gas media, in particular to those of dark matter search
\cite{Zeplin,Xenon,Warp}, $\mu$-e-$\gamma$ decay search \cite{Meg},
neutrinoless double beta decay search \cite{Next} and coherent
neutrino-nucleus scattering \cite{CoNu1,CoNu2}. For example, the
difference in light yields of primary scintillations produced by
nuclear recoils and $\gamma$-quanta in liquid Ar and Xe, provides an
effective means for selecting WIMP signals against the background in
dark matter search experiments using two-phase noble gas detectors
\cite{NobleRev}, i.e. in ZEPLIN \cite{Zeplin}, XENON \cite{Xenon}
and WARP \cite{Warp} experiments. In addition, to record the
ionization signal along with that of the primary scintillation,
secondary (proportional) scintillations are used in such
experiments, i.e. those of electroluminescence generated by
electrons drifting under moderate electric field in the gas phase
\cite{NobleBook,SecScint}. The other examples of noble gas
scintillation applications are the noble liquid scintillation
calorimetry, in particular that of the MEG experiment \cite{Meg}
based on recording primary scintillations in liquid Xe, and the
high-pressure noble-gas TPC of the NEXT experiment \cite{Next} based
on recording proportional scintillations in compressed Xe.

So far, noble gas scintillations in high energy physics experiments
have been recorded essentially in the vacuum ultraviolet (VUV). This
necessitated the use of sophisticated VUV-sensitive photodetectors,
including those with Wavelength Shifter (WLS) coatings. Indeed, in
the VUV the primary scintillation yield is rather high: of about
(40-60)$\times$10$^3$ photon/MeV in liquid Ar and Xe \cite{NobleRev}
and 14$\times$10$^3$ photon/MeV in gaseous Xe \cite{GXeScint}. The
VUV emission is caused by reactions between excited and ionized
atoms producing excimers which decay radiating the VUV continua
\cite{VUVScint}. In addition in condensed noble gases, a mechanism
of VUV luminescence due to excitation of excitons might play a role
\cite{VUVExciton}. At high pressures the VUV emission was generally
believed to dominate over all other types of radiative decays such
as atomic emission in the visible and infrared regions
\cite{NobleRev}.

On the other hand, as early as 20 years ago it was suggested that
this statement might not be valid due to the discovery of intense
atomic emission scintillations in practically all noble gases in the
near infrared (NIR) \cite{NIR1}: in particular in Ar in the
wavelength range of 690-850 nm \cite{NIR1} and in Xe at 800-1600 nm
\cite{NIR2}. Since then, the NIR emission spectra of scintillations
in gaseous and liquid Ar and Xe have been further studied in several
works: both for primary \cite{NIR2,NIR3,NIR6} and secondary
\cite{NIR4,NIR5} scintillations. In gaseous Ar this kind of
scintillation was attributed to transitions between the atomic
states of the Ar (3p$^5$ 4p) and Ar (3p$^5$ 4s) configurations
\cite{NIR1,NIR4}. However, little is known about the absolute NIR
scintillation yield in noble gases: almost nothing about that in Ar
and only the lower limit in gaseous Xe ($\geq$21$\times$10$^3$
photon/MeV) \cite{NIR5}.

In this paper we present the experimental data in support for the
hypothesis of the intense NIR scintillations in Ar in view of its
potential application in rare-event experiments: for the first time
we have measured the scintillation yield in gaseous and liquid Ar in
the NIR and visible region, using Geiger-mode Avalanche Photodiodes
(G-APDs, \cite{GAPDRev}). The latter have rather high photon
detection efficiency (PDE) in the wavelength range of 400-1000 nm,
of about 15\% on average \cite{CPTA}, providing direct and effective
detection of NIR scintillations without WLS.

This work has actually been triggered by the recent results on NIR
scintillations obtained in our laboratory \cite{OptRead1}: a rather
high avalanche (secondary) scintillation yield, of about 4 NIR
photons per avalanche electron, was observed. This observation was
done in the course of the development of a novel technique of signal
recording in detectors of ultimate sensitivity for rare-event
experiments, such as those of two-phase Ar avalanche detectors
\cite{TPArTHGEM} based on thick Gas Electron Multipliers (THGEMs)
\cite{THGEM} with optical readout using G-APDs
\cite{OptRead1,OptRead2}. We will discuss this technique in the
following, as well as other possible applications of NIR
scintillations in high energy physics experiments.

\section{Experimental setup}

\begin{figure}
\includegraphics[width=0.8\columnwidth]{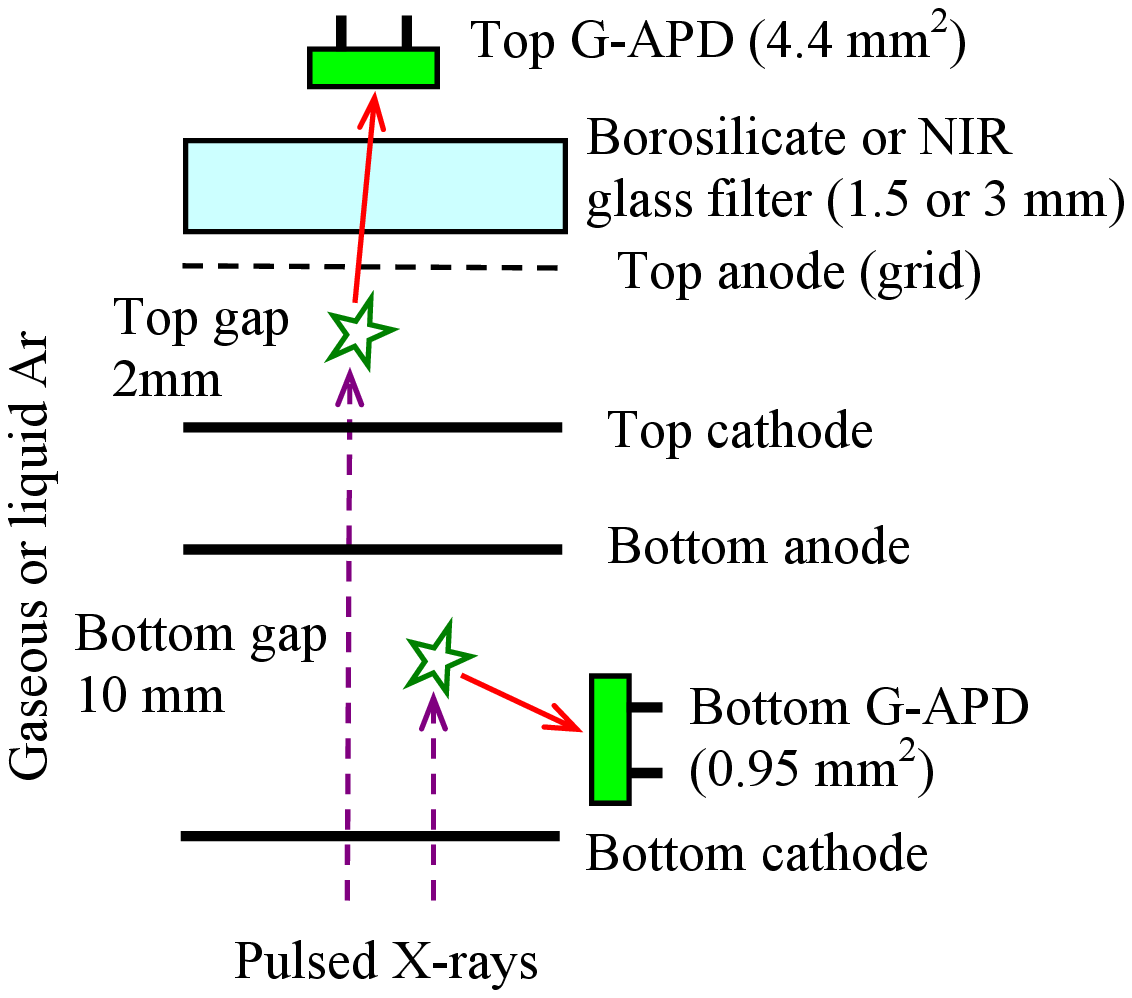}
\caption{Schematic view of the experimental setup to study
scintillations in the NIR and visible region in gaseous and liquid
Ar (not to scale).} \label{setup}
\end{figure}

Fig.~\ref{setup} shows the experimental setup; as concerns
cryogenics and electronics it was similar to that described
elsewhere \cite{OptRead1}. It consisted of a 9 l cryogenic chamber
filled with either gaseous or liquid Ar. Ar was taken from the
bottle with a specified purity of 99.998\% (N$_2$ content $<$10ppm);
during cooling procedures it was additionally purified from oxygen
and water by Oxisorb filter \cite{Oxisorb}, providing electron
life-time in the liquid $>$20  $\mu$s \cite{TPArGEM} which
corresponds to the oxygen-equivalent impurity content below 20 ppb.

To cross-check the correctness of the scintillation yield
measurement procedure, several measurement runs were conducted in a
variety of ways: using two different photodetectors, at different
solid angles, with and without NIR transmission optical filter. In
particular, the scintillation and ionization signals were recorded
in two different gaps, top and bottom, with an active area of
30$\times$30 mm$^2$ and thickness of 2 and 10 mm respectively (see
Fig.1). The gaps were formed by thin-plate electrodes made of
copper-clad G10 and a wire grid as shown in Fig.~\ref{setup}.

The signals in the gaps were induced by X-rays from a pulsed X-ray
tube with a Mo anode operated at a voltage of 40 kV. The gaps were
irradiated from outside the cryogenic chamber, practically uniformly
across the active area, through a lead collimator and aluminium
windows at the chamber bottom, defining a cylindrical X-ray
conversion region of a 20 mm diameter in each gap. Accordingly, the
energies of incident X-rays were within 15-40 keV in gaseous Ar and
30-40 keV in liquid Ar. The X-ray pulse had a sufficient power to
provide measurable ionization charges in the gaps, typically having
values of several thousands and tens of thousands electrons in
gaseous and liquid Ar respectively. In addition it was sufficiently
fast, having a width of 0.5  $\mu$s, to provide a reasonable
scintillation time resolution.

The gaps were viewed by respectively top and bottom G-APDs as shown
in Fig.~\ref{setup} (manufactured by CPTA company \cite{CPTA}),
namely by "CPTA 149-35" and "CPTA 143-22" having 4.41 and 0.95
mm$^2$ sensitive area, respectively. The G-APDs were sensitive in
the wavelength range of 400-1000 nm: at the gaseous Ar emission
spectrum maximum (800 nm \cite{NIR1,NIR4}) they had PDE$\sim$15\%
\cite{CPTA}. In front of the top G-APD either a borosilicate glass
filter, transparent to visible and NIR light, or NIR glass filter,
transparent to NIR above 700 nm, was placed. The bottom G-APD had no
optical filter. It is essential that the measurements were carried
out at cryogenic temperatures, in the range of 87-163 K, where
G-APDs have a superior performance as compared to room temperature,
in terms of the noise rate, amplitude and single-photoelectron
characteristics \cite{CryoGAPD}.

The scintillation signal was read out from either the top or bottom
G-APD using a fast amplifier (CPTA, \cite{CPTA}) with 300 MHz
bandwidth. The charge (ionization) signal was read out from the
anode electrode of either the top or bottom gap using a
charge-sensitive amplifier. Both amplifiers were placed outside the
cryogenic chamber. The signals were digitized and memorized for
further off-line analysis with a TDS5032B digital oscilloscope.

\section{Measurement procedure}

It should be remarked that any measurement of the absolute
scintillation yield is generally a difficult task, since one has to
correctly determine the number of photons emitted in the
scintillation burst. In the present work we applied a new approach
to make such measurements, namely using the G-APD in single
photoelectron counting mode with time resolution. As compared to
other photodetectors, in addition to its higher sensitivity in the
NIR, the G-APD provides a faster single-photoelectron pulse of a
well-defined shape, with a width of 20 ns \cite{OptRead1,CryoGAPD}.
These resulted in a remarkable capability of the G-APD to count the
number of photoelectrons contained in a given (digitized)
scintillation signal: by counting the number of single-photoelectron
pulses in a time scale corresponding to the scintillation signal
duration. To realize this method, the X-ray pulse intensity was
reduced down to a level providing a practically non-overlapping of
single-photoelectron pulses in the scintillation signal. This was
the key element of the measurement procedure; the other details of
the procedure were as follows.

The scintillation yield is defined as the ratio of the number of
photons emitted over full solid angle ($N_{ph}$) to the number of
ionization electrons created in the gap ($N_e$). In practice Y was
obtained from the G-APD's photoelectron yield over full solid angle
($Y_{pe}$) and the PDE averaged over the spectrum of the Ar
emission passed through the optical filter ($<PDE>$): \\
$Y = N_{ph} / N_e  = Y_{pe} / <PDE>$ .   \hfill               (1) \\
$<PDE>$ was calculated using a G-APD's PDE spectrum provided by the
manufacturer \cite{CPTA} and Ar emission spectrum taken from Refs.
\cite{NIR1} and \cite{NIR6} for gaseous and liquid Ar respectively:
$<PDE>$ was equal to 17.6\% and 16.7\% respectively when using the
borosilicate glass filter and to 12.6\% when using the NIR filter.

$Y_{pe}$ was obtained from the charge signal amplitude ($N_e$),
scintillation signal amplitude ($N_{pe}$) and G-APD average
acceptance with respect to the X-ray conversion region
($A_{G-APD}$), equal to the average reduced solid angle
($<\Delta\Omega_{G-APD}/4\pi>$): \\
 $Y_{pe} = N_{pe} / N_e/ A_{G-APD}$  . \hfill                 (2)

$A_{G-APD}$ was accurately calculated using Monte-Carlo simulation.
In these calculations one took into account the light reflection
from the electrodes for both gaps; in addition for the top gap, the
light refraction, Frenel reflections and absorption in the optical
filter, if any, as well as the anode grid transparency were taken
into account. The calculated $A_{G-APD}$ values amounted to 0.0029
for the top G-APD and 0.00018 for the bottom G-APD. Notice a
considerable difference, of a more than an order of magnitude, in
acceptances for the top and bottom G-APDs.

$N_e$ is just the ionization signal amplitude expressed in
electrons; it was measured at the anode electrode under electric
field applied across the gap. Similarly, $N_{pe}$ is the
scintillation signal amplitude expressed in photoelectrons; it was
measured using the method described above.

Regarding the measurement errors, the major sources of uncertainties
were classified as follows: the $N_e$ and $N_{pe}$ measurements gave
10-30\% and 10-20\% respectively (depending on a particular
measurement run), the $<PDE>$ calculation gave 5\%. In total these
resulted in about 15-40\% uncertainty in the scintillation yield.

Further details of the experimental setup and procedure will be
presented in our more elaborated paper \cite{NextPaper}.

\section{Results and discussion}

In this paper we present the results obtained at cryogenic
temperatures: in gaseous Ar - at a temperature of about 160 K and
pressure of 0.60 atm (with gas density corresponding to that of 1
atm at room temperature) and at 87 K and 1.0 atm (i.e. in saturated
vapor); in liquid Ar - at 87 K and 1.0 atm.  The scintillations in
the G-APD sensitivity region were observed under these conditions
and their yields were measured in both gaseous and liquid Ar.

The scintillation signal had a fast component with a time structure
reflecting that of the X-ray pulse, thus having a time constant
below 0.5  $\mu$s. This time scale was expected from the atomic
emission scintillation mechanism. On the other hand in addition to
the fast component, a slow scintillation component was observed in
both gaseous and liquid Ar, with a time constants of about 20
$\mu$s. Its integral contribution to the light yield, in a ~50
$\mu$s time interval, was comparable to that of the fast component.
The origin of the slow component has not been yet understood. Most
probably it is caused by slow scintillations of Ar or those of
impurities of yet unclear nature; other hypotheses such as delayed
X-ray fluorescence, optical filter fluorescence and G-APD
after-pulses were tested and practically excluded in supplementary
measurements. Accordingly in the following, only the fast component
contribution to the scintillation yield will be taken into
consideration; the slow component contribution was accurately
measured and subtracted.

\begin{figure}
\includegraphics[width=0.99\columnwidth]{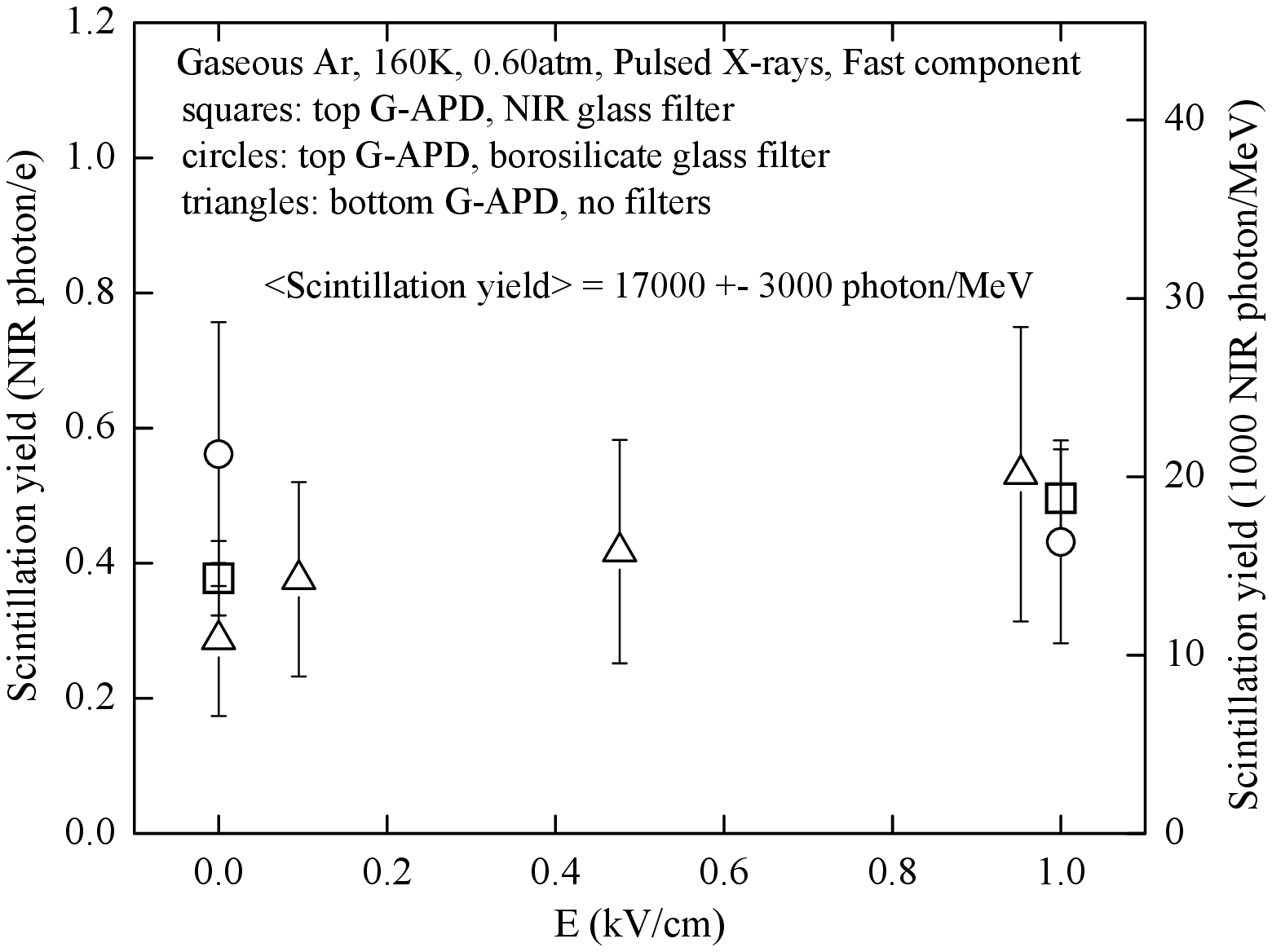}
\caption{Primary NIR scintillation yield for the fast component in
gaseous Ar at 160 K and 0.60 atm, measured in the electric field
range of 0-1 kV/cm under different conditions: with top G-APD and
NIR glass filter, with top G-APD and borosilicate glass filter, with
bottom G-APD and no optical filter. The scintillation yield is
expressed in the number of photons per primary ionization electron
(left scale) and photons per MeV of deposited energy (right scale)}
\label{GArYield}
\end{figure}

In gaseous Ar, at low and moderate electric fields applied across
the gap, varying from 0.1 to 4 kV/cm, the gaps were operated in an
ionization mode. i.e. without electron multiplication. At these
fields the collected charge was practically independent of the
field, enabling to unambiguously determine the primary ionization
charge ($N_e$).  The primary scintillation yield was measured at low
electric fields, i.e. in the absence of secondary scintillations, in
the range of 0-1 kV/cm: it is shown in Fig.~\ref{GArYield}. Three
groups of data are presented obtained under substantially different
conditions: with the top G-APD and NIR glass filter, with the top
G-APD and borosilicate glass filter, with the bottom G-APD and no
optical filter.

One can see that there is a good agreement between all the data
within their accuracies. Hence one may conclude the following.
Firstly,  the scintillations took place essentially in the NIR,
since the scintillation yields obtained with and without the NIR
filter were close to each other. Secondly, the NIR emission spectrum
in our measurements is similar to that of Ref. \cite{NIR1}, since
just this spectrum provided the data compatibility. Thirdly, the
acceptances were calculated correctly, since the data obtained in
the top and bottom gaps are consistent, despite the considerable
difference in measurement geometries and G-APD solid angles.

The primary NIR scintillation yield in gaseous Ar, averaged over all
the measurements presented in Fig.~\ref{GArYield}, is $Y_{GAr} =
0.44 \pm 0.09$ photon/e. It is convenient to express the yield in
the number of photons per MeV of deposited energy considering a
W-value (energy needed to produce one ion pair) in gaseous Ar of
26.4 eV \cite{NobleRev}: $Y_{GAr} = 17000 \pm 3000$ photon/MeV. It
should be remarked that this primary scintillation yield was found
to be independent of the temperature in the range of 87-163 K. It is
also remarkable that this yield value is close to that of the VUV
for gaseous Xe \cite{GXeScint} and to that of the low limit
established for NIR scintillations in Xe \cite{NIR5}. Thus, one may
conclude that the primary scintillation yield is rather similar for
all noble gases, both in the VUV and NIR regions.

\begin{figure}
\includegraphics[width=0.99\columnwidth]{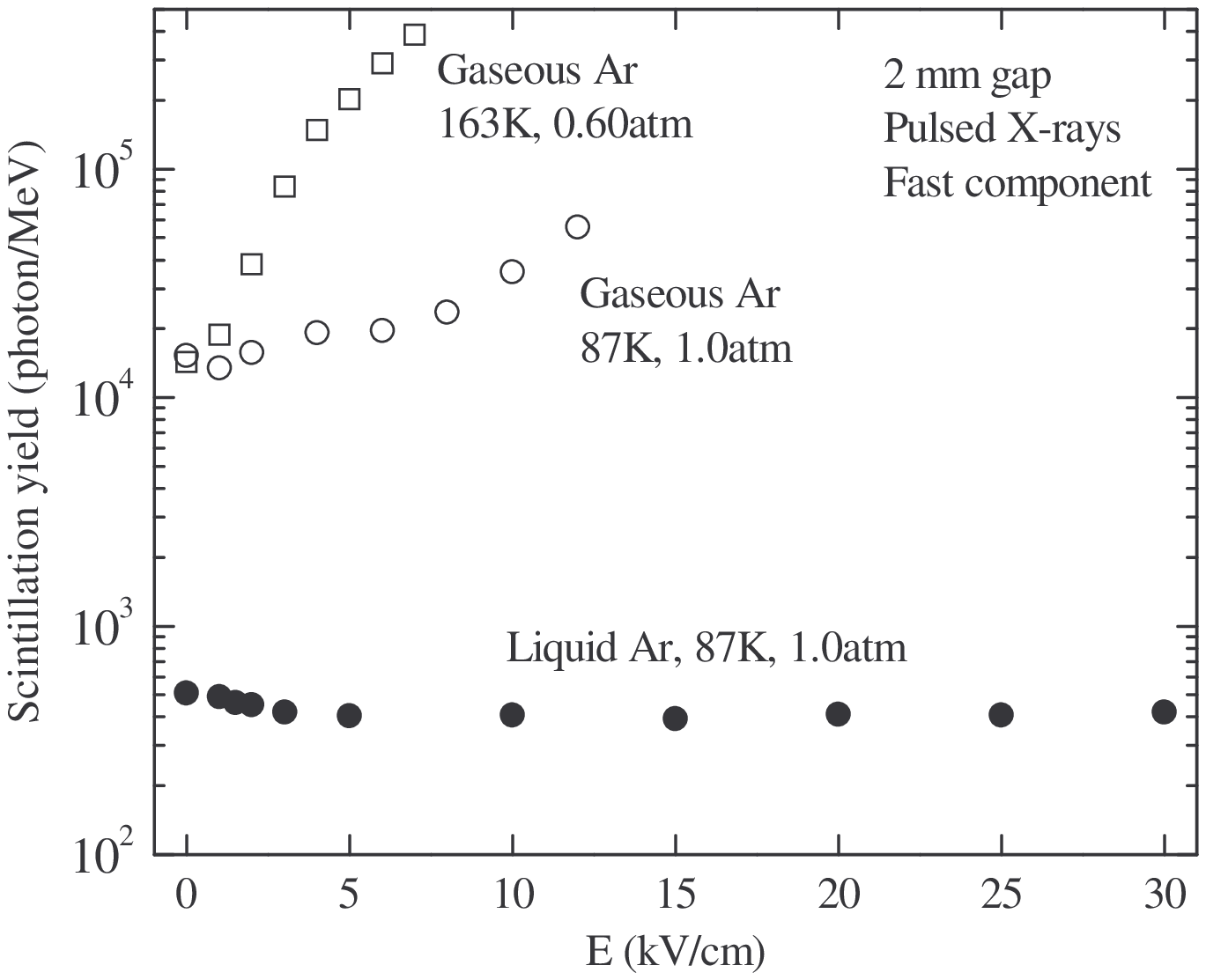}
\caption{Scintillation yield for the fast component as a function of
the electric field in liquid Ar in the NIR and visible region, at 87
K and 1.0 atm, and in gaseous Ar in the NIR, at 163 K and 0.60 atm
and at 87 K and 1.0 atm.  The scintillation yield is given in the
number of photons per MeV of deposited energy of the primary
ionization. The data were obtained under X-ray irradiation in a 2 mm
thick gap.}
\label{LArYield}
\end{figure}

In liquid Ar, the charge collected in the gap ($N_c$) depends on the
electric field ($E$) due to the recombination effect. In our
measurements this dependence was described in the frame of the
recombination model of Ref. \cite{NobleBook} enabling to determine
the primary ionization charge $(N_e)$: $N_c = N_e / ( 1 + k / E )$.
In liquid Ar the non-VUV emission spectrum was reported to be
continuous \cite{NIR6}, extending from 400 to 1000 nm, in contrast
to the atomic line spectrum in gaseous Ar. Accordingly, for liquid
Ar we used the borosilicate glass filter transparent in both the NIR
and visible range. Taking these facts into account,
Fig.~\ref{LArYield} shows the scintillation yield in liquid Ar at 87
K, in the NIR and visible region, as a function of the electric
field varying from 0 to 30 kV/cm. For comparison, the NIR
scintillation yield in gaseous Ar is also shown at the same
temperature, i.e. in saturated vapor at 87 K and 1.0 atm, as well as
at 163 K and 0.60 atm. The scintillation yield in
Fig.~\ref{LArYield} is given in the number of photons per MeV of
deposited energy of the primary ionization, using a W-value of 26.4
and 23.6 eV for gaseous and liquid Ar respectively \cite{NobleRev}.
The primary scintillation yield in liquid Ar, defined at zero field,
amounts to $Y_{LAr} = 510 \pm 90$ photon/MeV. Notice that it is
considerably reduced, by a factor of 30, compared to that of gaseous
Ar. The effect of suppression of NIR scintillations in the noble
liquid as compared to that of the noble gas was observed earlier for
Xe \cite{NIR2}. The mechanism of such suppression is yet unclear.

From Fig.~\ref{LArYield} one can see that the scintillation yield in
gaseous Ar increased with the electric field starting from a certain
threshold: in particular in Ar at 163 K and 0.60 atm - from 2 kV/cm.
This increase is explained by secondary scintillations
(electroluminescence) induced by electrons drifting and exciting
atoms under moderate electric fields, at which the gap is operated
in an ionization mode, i.e. without electron multiplication. This
kind of scintillations is called proportional scintillations due to
the proportionality of their intensity to the electric field. The
effect of proportionality of NIR scintillations in Ar was observed
earlier at room temperature \cite{NIR3}, though their absolute yield
was not measured. In the present work, in gaseous Ar at 163 K the
law of proportionality was valid up to 6 kV/cm. At this field the
scintillation yield increased by an order of magnitude compared to
that of the primary scintillations, reaching a value of
200$\times$10$^3$ photon/MeV (or 5 photon/e) for the effective gap
thickness of 1 mm. Consequently, we confirm here the high secondary
NIR scintillation yield observed recently in our laboratory using
THGEM multiplier \cite{OptRead1}: at avalanche gains of 60-400 it
was of the order of 4 photon/e.

In denser gaseous Ar, at 87 K and 1 atm, proportional scintillations
were also observed, though at higher threshold (above 8 kV/cm). In
liquid Ar in contrast, no proportional scintillations were observed
at all, even at fields reaching 30 kV/cm (see Fig.~\ref{LArYield}).

\begin{figure}
\includegraphics[width=0.99\columnwidth]{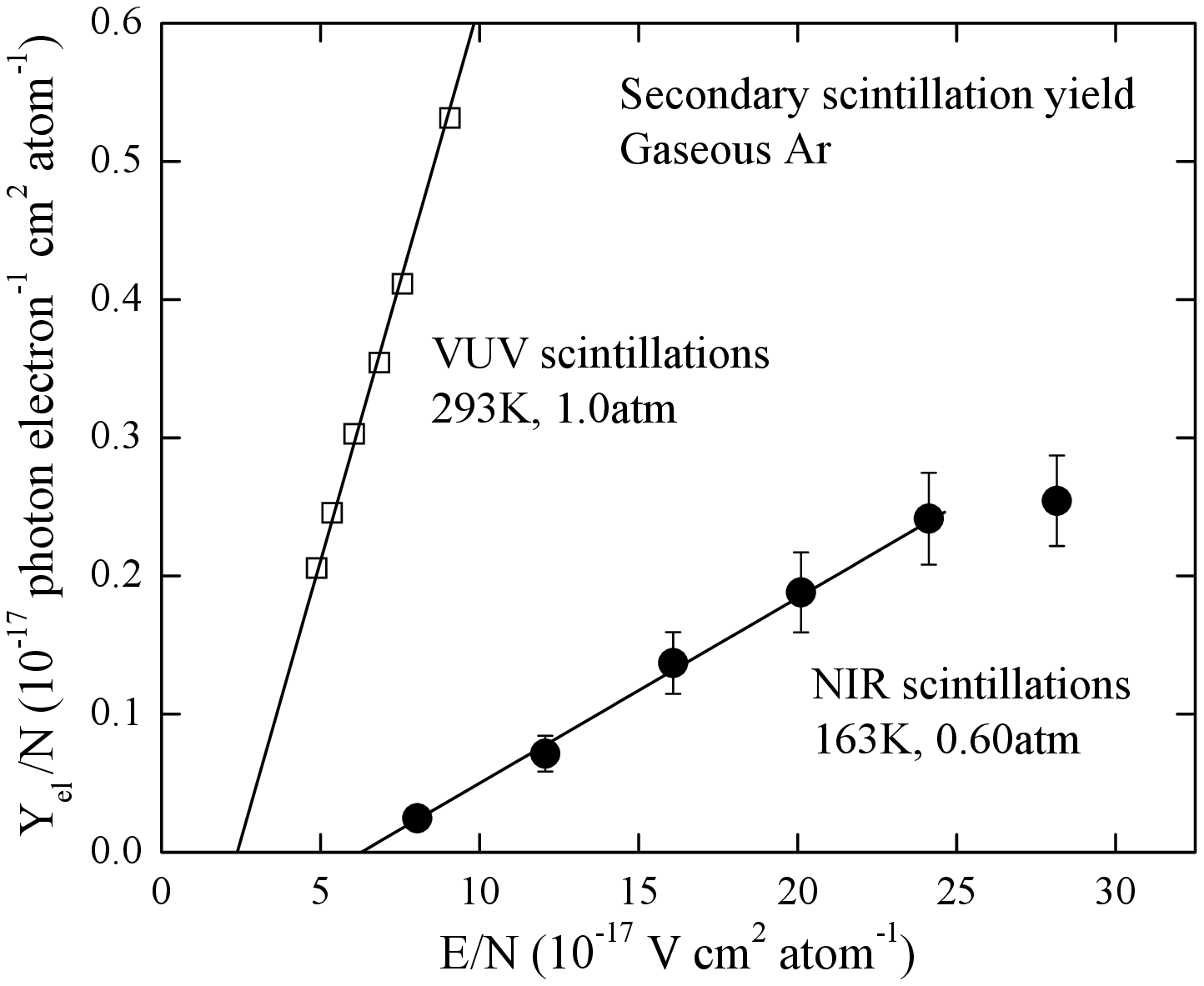}
\caption{Reduced NIR electroluminescence yield for the fast
component as a function of the reduced electric field in gaseous Ar
at 163 K and 0.60 atm measured in the present work; for comparison
that of the VUV measured elsewhere \cite{SecScint} is shown.}
\label{GArSecScint}
\end{figure}

The electroluminescence yield is defined as the number of photons
normalized to the total ionization charge generated in the gap, the
primary scintillation contribution being subtracted, and to the
average electron drift path in the gap ($d$), equal to 1 mm for the
top gap: \\
$Y_{el} = N_{ph} / N_e / d$ . \hfill              (3)\\
To compare our electroluminescence results obtained in the NIR to
those obtained in the VUV, the reduced electroluminescence
scintillation yield is shown in Fig.~\ref{GArSecScint} as a function
of the reduced electric field. The linear part of the yield
dependence corresponds to proportional scintillations; in the NIR at
163 K it is described by the following equation: \\
$Y_{el}/N$ [10$^{-17}$ photon electron$^{-1}$ cm$^2$ atom$^{-1}$]$ =
0.013E/N - 0.084$ . \hfill (4)\\
Here $N$ is the atomic concentration and $E/N$ is given in Td
[10$^{-17}$ V cm$^2$ atom$^{-1}$]. This equation is universally
valid for any temperature and pressure. Thus it may be compared to
the proportional scintillation yield in the VUV at room temperature
presented elsewhere \cite{SecScint}: \\
$Y_{el}/N$ [10$^{-17}$ photon electron$^{-1}$ cm$^2$ atom$^{-1}$]$ =
0.081E/N - 0.190$ . \hfill (5)

One can see that the electroluminescence amplification parameter in
the NIR, defined as the number of photons produced per drifting
electron and per kilovolt, equivalent to the line slope in
Fig.~\ref{GArSecScint}, is equal to 13; this is by a factor of 6
lower compared to that of the VUV. One can also see that
proportional scintillations exist in the NIR up to reduced field
values of 25 Td. At higher fields the gap started to operate in a
proportional mode, i.e. with electron multiplication, at which the
atomic excitation processes are taken over by those of the avalanche
ionization, resulting in the reduction of the electroluminescence
efficiency.

\begin{figure}
\includegraphics[width=0.8\columnwidth]{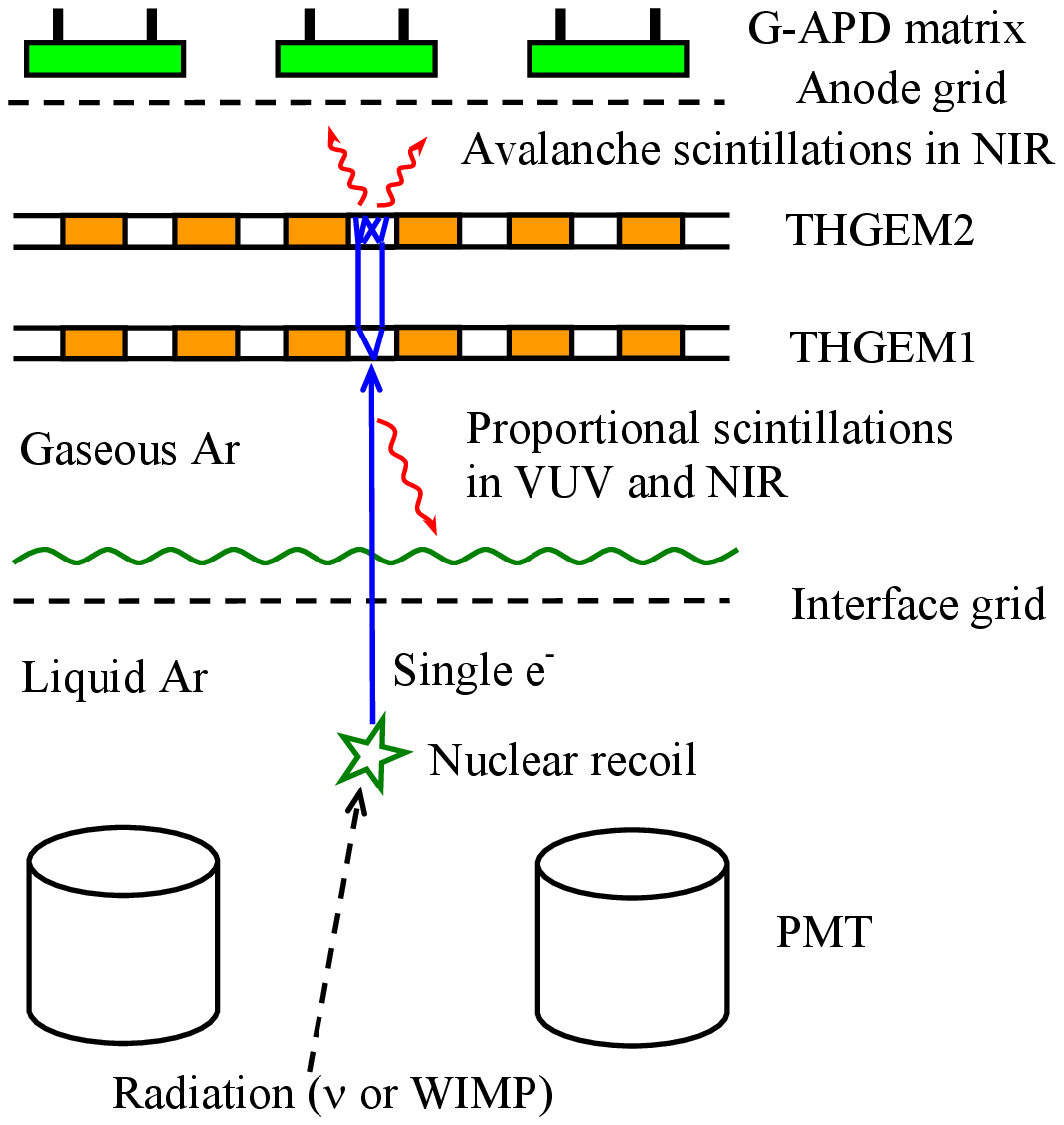}
\caption{Principles of operation of a THGEM-based two-phase Ar
avalanche detector with G-APD optical readout for coherent
neutrino-nucleus scattering and dark matter search experiments.}
\label{OptRead}
\end{figure}

Let us consider now possible applications of NIR scintillations. As
mentioned in the Introduction, the first application is the
THGEM-based two-phase Ar avalanche detector of ultimate sensitivity
(i.e. operating in single-electron counting mode at extremely low
noise) with G-APD optical readout \cite{OptRead1}, for coherent
neutrino-nucleus scattering and dark matter search experiments. The
principles of operation of such a detector are shown in
Fig.~\ref{OptRead}; these are the further development of the concept
suggested in \cite{CoNu2} for coherent neutrino-nucleus scattering
experiment. The ionization, produced by nuclear recoil from a weakly
interacting particle, is extracted from the liquid into the gas
phase, where it is amplified in two stages: first, in a proportional
scintillation mode in a gap above the liquid-gas interface, and
second, in an avalanche scintillation mode in a THGEM multiplier.
Proportional scintillations are recorded in the VUV and NIR using
PMTs placed in the liquid: this will provide a single-photoelectron
trigger due to the excellent amplitude resolution available in the
proportional scintillation mode. Avalanche scintillations are
recorded in the NIR using a matrix of bare G-APDs: this will provide
a good spatial resolution. Such an optical readout is preferable as
compared to charge readout in terms of overall gain and noise. In
particular it has been demonstrated \cite{OptRead1} that the
THGEM/G-APD combined multiplier could operate in a single electron
counting mode in a two-phase Ar avalanche detector, at an avalanche
gain of 400. A practical detector would comprise a G-APDs matrix
placed a few millimeters behind the THGEM multiplier, with a pitch
of ~1 cm, viewing clusters of multiplier holes under an angle of
$\pm$45$^{\circ}$. For example, for a 100 kg liquid Ar TPC of a
volume of 40$\times$40$\times$40 cm$^3$ the total number of G-APDs
would be reasonable - of 1600. Such a detector would be robust,
stable, simple and relatively cheap.

The other application might be noble liquid non-VUV scintillation
calorimetry. The scintillation yield in liquid Ar measured here in
the NIR and visible region ($\sim$500 photon/MeV) might be enough
for high energy calorimetry: it is comparable with that of the fast
solid scintillators being already used in calorimeters, namely
higher than that of PbWO$_4$ ($\sim$100 photon/MeV) and somewhat
lower than that of pure CsI ($\sim$2000 photon/MeV).  The NIR
scintillation yield in liquid Xe is expected to be of the same order
as that of Ar, providing the applicability in liquid Xe NIR
scintillation calorimetry. The readout of the calorimeters might be
performed using NIR-sensitive photodetectors: silicon APDs and
G-APDs for liquid Ar and InGaAs photodiodes for liquid Xe.

One more application is the project for directional solar neutrino
detection using a GEM-based cryogenic Ne or He detector with CCD
optical readout (so-called "e-bubble" project) \cite{Ebubble}. Here
the property of intense NIR scintillations in noble gases is
combined with that of the high CCD sensitivity in the NIR.

\section{Conclusion}

The scintillation yield in gaseous and liquid Ar has for the first
time been measured in the NIR and visible region, both for primary
and secondary (proportional) scintillations, using Geiger-mode
avalanche photodiodes (G-APDs) and pulsed X-ray irradiation.

In gaseous Ar, the non-VUV scintillations took place essentially in
the NIR: the primary scintillation yield of the fast component was
measured to be $17000 \pm 3000$ photon/MeV or $0.44 \pm 0.09$
photon/e, in the range of 690-1000 nm. This is comparable with that
of the VUV for gaseous Ar and Xe.

In liquid Ar, the primary scintillation yield of the fast component
was considerably reduced, amounting to $510 \pm 90$ photon/MeV in
the range of 400-1000 nm. Nevertheless this is comparable with the
yields of fast solid scintillators used in high-energy calorimetery.

Proportional scintillations (electroluminescence) in the NIR have
been observed in gaseous Ar; their amplification parameter at 163 K
was measured to be 13 photons per drifting electrons per kV. Though
having somewhat lower yield than that in the VUV, proportional
scintillations may substantially increase the scintillation yield as
compared to that of primary scintillations in the NIR: by an order
of magnitude, to hundreds of thousands photons per MeV.

No proportional scintillations have been observed in liquid Ar up to
the electric fields of 30 kV/cm.

There is potentially a wide variety of applications of noble gas NIR
scintillations in high energy physics experiments. Among them are
THGEM-based two-phase Ar avalanche detectors of ultimate sensitivity
with G-APD optical readout for coherent neutrino-nucleus scattering
and dark matter search experiments, liquid noble scintillation
calorimetry, GEM-based cryogenic detectors with CCD optical readout
for solar neutrino detection, and others.

\acknowledgments
We are grateful to Y. Tikhonov for the support and
D. Akimov for discussions. This work was supported in part by
Federal special program "Scientific and scientific-pedagogical
personnel of innovative Russia" in 2009-2013.

\end{document}